\documentclass[aps,prd,preprintnumbers,twocolumn,superscriptaddress,nofootinbib]{revtex4}
\usepackage{hyperref}
\hypersetup{colorlinks=true,linkcolor=purple,anchorcolor=blue,citecolor=blue, filecolor=blue,urlcolor=red,bookmarksnumbered=true,
	pdfview=FitB
}
\usepackage{color}
\usepackage{xcolor}
\usepackage{ulem}
\usepackage{csquotes}
\colorlet{purple1}{blue!70!red}
\colorlet{darkred}{red!50!black}

\usepackage{graphicx}
\usepackage[bf,SL,BF]{subfigure}
\usepackage{psfrag}
\usepackage{color}
\usepackage{amssymb}
\usepackage{amsmath}
\usepackage{epstopdf}
\usepackage{natbib}
\usepackage[mathlines]{lineno}

\def\orcid#1{\kern .08em\href{https://orcid.org/#1}{\includegraphics[keepaspectratio,width=0.7em]{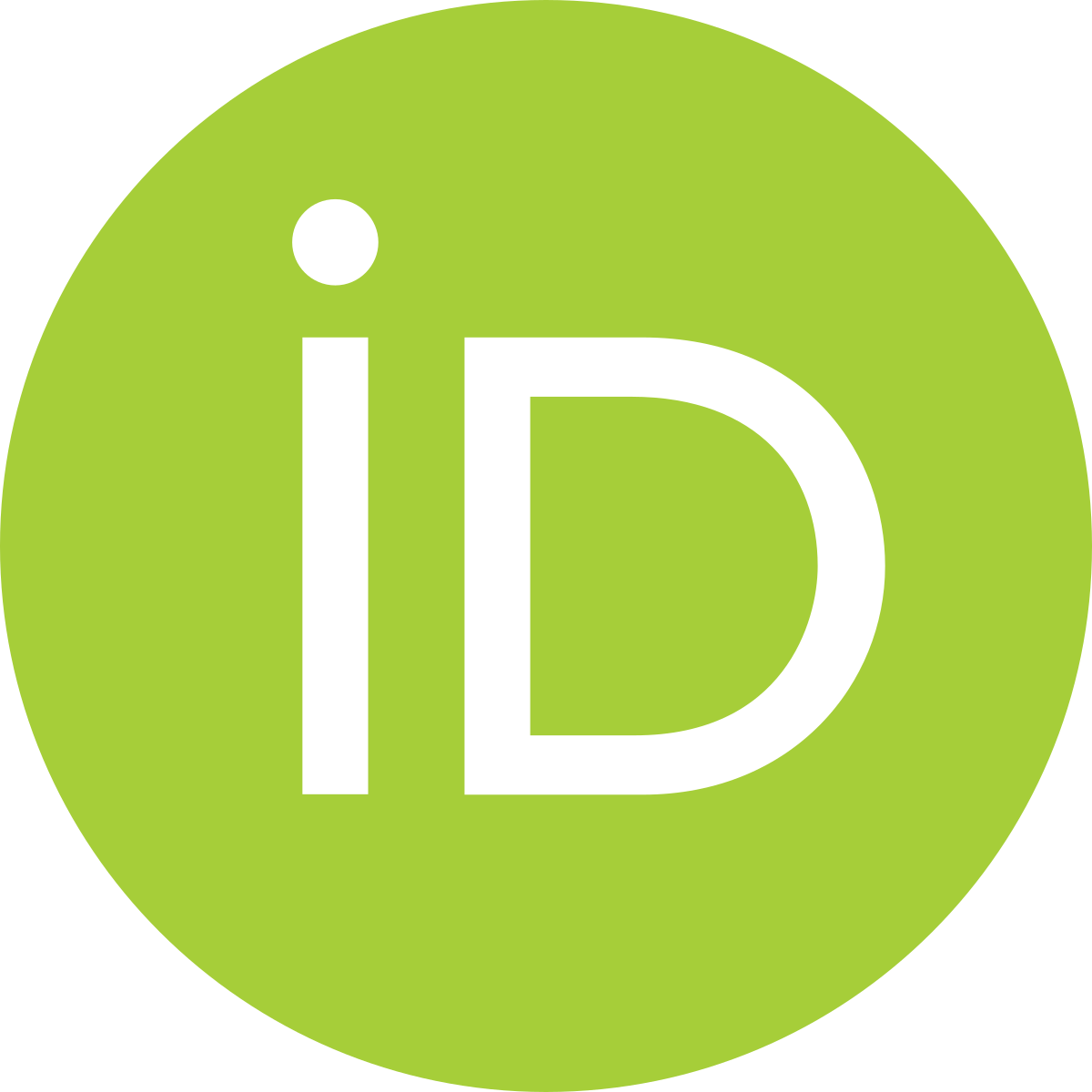}}}
\newcommand{\be}{\begin{eqnarray}}
	\newcommand{\ee}{\end{eqnarray}}
	
\def\orcid#1{\kern .08em\href{https://orcid.org/#1}{\includegraphics[keepaspectratio,width=0.7em]{ORCID_iD.png}}}

\newcommand{\kp}{{k}_{\perp}}

\newcommand{\bp}{{b}_{\perp}}

\begin{document}

	\title{Predicting $\sin(2\phi-\phi_{s})$ azimuthal asymmetry in pion-proton induced Drell-Yan process using holographic light-front QCD}	
	
	\author{Bheemsehan~Gurjar\orcid{0000-0001-7388-3455}}
	\email{gbheem@iitk.ac.in} 
	\affiliation{Indian Institute of Technology Kanpur, Kanpur-208016, India}
	
	\author{Chandan~Mondal\orcid{0000-0002-0000-5317}}
		\email{mondal@impcas.ac.cn} 
	\affiliation{Institute of Modern Physics, Chinese Academy of Sciences, Lanzhou 730000, China}
	\affiliation{School of Nuclear Science and Technology, University of Chinese Academy of Sciences, Beijing 100049, China}

	\date{\today}
	
	\begin{abstract}
		We compute the $\sin(2\phi-\phi_{s})$ azimuthal asymmetry in the pion-nucleon induced Drell-Yan process within  transverse momentum dependent factorization. We employ the holographic light-front pion wave functions to calculate its leading-twist transverse momentum dependent parton distributions (TMDs). The Boer-Mulders TMD of the pion is then convoluted with the transversity TMD of the proton evaluated in a light-front quark-diquark model constructed with the wave functions predicted by the soft-wall
AdS/QCD to obtain the azimuthal asymmetry in the Drell-Yan process. The gluon rescattering is pivotal to predict nonzero pion Boer-Mulders TMD. We investigate the utility of a
nonperturbative SU$(3)$ gluon rescattering kernel going beyond the usual approximation of perturbative U$(1)$ gluons. The holographic light-front QCD approach provides a powerful tool for exploring the role of nonperturbative QCD effects in the Drell-Yan process and may help to guide future experimental measurements. 	
	\end{abstract}

	\maketitle
	
	\section{Introduction}

	Transverse momentum dependent distribution functions (TMDs)~\cite{Collins:1981uw, Collins:2003fm, Collins:2011zzd,Tangerman:1994eh, Kotzinian:1994dv, Mulders:1995dh} give novel insights into three-dimensional (3D) partonic structure of hadrons by accounting for parton transverse motion and spin-orbit correlations. At leading-twist, there exist two quark TMDs for a spin-$0$ hadron~\cite{Meissner:2008ay}, while for a spin-$1/2$ hadron, there are eight twist-2  quark TMDs~\cite{Meissner:2009ww, Meissner:2007rx}. One of them is the Boer-Mulders function, denoted as $h_{1}^{\perp}(x,k^2)$~\cite{Boer:1999mm,Boer:1997nt}. It shows the connection between the quark spin and the quark transverse momentum, which leads to the transversely polarized asymmetries of quark inside an unpolarized hadron. However, the existence of the Boer-Mulders function was not so obvious initially. Under (naive) time reversal invariance of QCD~\cite{Collins:1992kk}, the Boer-Mulder function was considered to vanish with its chiral even partner so called, Sivers function~\cite{Sivers:1989cc}. Explicit model calculations~\cite{Brodsky:2002cx, Brodsky:2002rv, Boer:2002ju} incorporating gluon exchange between the struck quark and the spectator system indicate that the T-odd distributions can actually survive through the Wilson lines~\cite{Collins:2002kn, Ji:2002aa}. The presence of Wilson line also shows the process dependence of the T-odd, Sivers and Boer-Mulders functions, i.e., they flip signs between the semi-inclusive deeply inelastic scattering (SIDIS) and the Drell-Yan processes~\cite{Brodsky:2002rv, Boer:2002ju, Collins:2002kn}, which is a crucial prediction that will need to be confirmed by future experiments. In recent decades, various QCD inspired models and phenomenological analyses have been used extensively  to study the Boer-Mulders function of the proton and the pion~\cite{Boer:2002ju, Gamberg:2003ey, Yuan:2003wk, Pobylitsa:2003ty, Bacchetta:2003rz, Lu:2004au, Lu:2006ew, Gamberg:2007wm, Burkardt:2007xm, Bacchetta:2008af, Zhang:2008nu, Meissner:2008ay, Courtoy:2009pc, Gamberg:2009uk, Lu:2009ip, Barone:2009hw,Barone:2010gk, Pasquini:2010af, Lu:2012hh, Pasquini:2014ppa, Lu:2016pdp, Wang:2017onm, Gurjar:2022rcl, Gurjar:2021dyv, Maji:2017wwd}. 
	
	The Boer-Mulders function is a chiral-odd distribution function, hence in order to survive in a high energy scattering process, it must couple with another chiral-odd distribution/fragmentation function. The unpolarized Drell-Yan process, which exhibits an azimuthal dependency of the final-state dilepton with a $\cos2\phi$ modulation, is a promising method for obtaining the Boer-Mulders function. As proposed by Boer, such asymmetry can be generated by the coupling of two Boer-Mulder functions from each incoming hadron~\cite{Boer:1999mm}. The convolution of the Boer-Mulders function and the Collins fragmentation function $H_{1}^{\perp}$ in the unpolarized SIDIS process can give $\cos2\phi$ azimuthal asymmetry of the spin-0 produced hadron state. However, this asymmetry is tainted by the Cahn effect~\cite{Cahn:1989yf, Barone:2005kt, Barone:2008tn}, which is a higher-twist kinematical effect caused by the transverse motion of unpolarized quarks. The single transversely polarized Drell-Yan process provides another way to obtain the Boer-Mulders function. In this process the $\sin(2\phi-\phi_{s})$ asymmetry (with $\phi_{s}$, the azimuthal
	angle of target transverse spin) can be obtained through the convolution of the Boer-Mulders function $h_{1}^{\perp}(x,k^2)$ and the transversity distribution $h_{1}(x,k^2)$~\cite{Li:2019uhj,Bastami:2020asv}. Recently, the $\sin(2\phi-\phi_{s})$ asymmetry has been measured for the first time by the COMPASS experiment~\cite{COMPASS:2017jbv}, which has used a pion beam to collide with a transversely polarized nucleon target. Due to very high statistical uncertainty, the $\sin(2\phi-\phi_{s})$ asymmetry does not show a definite trend, although it does show a negative sign and a sizable magnitude.
	
	From the perspective of theory, several QCD inspired models such as the spectator model~\cite{Lu:2004au, Meissner:2008ay}, the light-front constituent quark model~\cite{Pasquini:2014ppa, Wang:2017onm, Wang:2018naw, Lorce:2016ugb}, the MIT bag model~\cite{Lu:2012hh}, and the Nambu--Jona-Lasinio model~\cite{Ceccopieri:2018nop, Noguera:2015iia}, etc., have made predictions for the nonzero pion Boer-Mulders function by using the perturbative U$(1)$ gluon rescattering. In Ref.~\cite{Gamberg:2009uk}, Gamberg and Schlegel have made a major attempt to overcome this perturbative approximation inside the antiquark spectator framework. The nonperturbative SU$(3)$ gluon rescattering kernel has been further employed to compute the nonzero Boer-Mulders TMD of the pion~\cite{Ahmady:2019yvo} and both the Sivers and the Boer-Mulders TMDs of the proton~\cite{Gurjar:2022rcl}.  
	
	In this work, we compute the $\sin(2\phi-\phi_{s})$ asymmetry of the pion-proton induced Drell-Yan process by considering the convolution $h_{1 (\pi)}^{\perp}\otimes h_{1(p)}$.  
We employ the Boer-Mulders function computed using a holographic light-front pion wave function with dynamical spin effects. These spin effects, together with gluon rescattering, are crucial to obtain a
nonzero holographic Boer-Mulders function~\cite{Ahmady:2019yvo}. Going beyond the usual approximation of perturbative U$(1)$ gluons, we investigate the use of a nonperturbative SU$(3)$ gluon rescattering kernel. On the other hand, we employ the transversity TMD of the proton evaluated in a light-front quark-diquark model constructed by the holographic light-front QCD.
We utilize the TMD factorization~\cite{Collins:1981uk, Collins:1984kg, Collins:2011zzd, Ji:2004xq}, which is appropriate when the transverse momentum of the dilepton $\vec{q}_\perp$ is much smaller than the hard scale $Q$, i.e., $\vec{q}_\perp\ll Q$. The TMD factorization has been extensively employed in the SIDIS~\cite{Collins:1981uk,Collins:2011zzd,Ji:2004wu,Aybat:2011zv,Collins:2012uy,Echevarria:2012pw}, $e^{+}e^{-}$ annihilation~\cite{Collins:2011zzd,Pitonyak:2013dsu,Boer:2008fr}, Drell-Yan~\cite{Collins:2011zzd,Arnold:2008kf} and W/Z boson production~\cite{Collins:2011zzd,Collins:1984kg,Lambertsen:2016wgj} processes. 
One of the important aspects of the TMD formalism is that it gives a systematic way to deal with the evolution of TMDs. In this TMD formalism, the scale evolution of TMDs is determined by the Collins-Soper equation~\cite{Collins:1981uk,Collins:1984kg,Collins:2011zzd,Idilbi:2004vb}. The solution of the evolution equation demonstrates that the changes of TMDs from one scale to another scale may be determined by an exponential form of the Sudakov-like form factor, which can be divided into perturbative and nonperturbative parts~\cite{Collins:1984kg,Collins:2011zzd,Aybat:2011zv,Collins:1999dz}. The perturbative Sudakov form factor is perturbatively calculable, while the nonperturbative Sudakov form factor is usually obtained by phenomenological extraction from experimental data. Here, we probe the scale evolution of the pion Boer-Mulders function as well as the proton transversity to estimate the $\sin(2\phi-\phi_{S})$ asymmetry at the COMPASS kinematics and compare our prediction with the latest COMPASS data~\cite{COMPASS:2017jbv} and the other theoretical predictions~\cite{Bastami:2020asv}.   
	
	The rest of the paper is organized as follows: In Sec.~\ref{pionTMDs}, we discuss the pion TMDs in the holographic light-front QCD framework whereas, the proton TMDs in a quark-diquark model motivated by the soft-wall AdS/QCD has been discussed in Sec.~\ref{ProtonTMDs}. In Sec.~\ref{TMDevolsec}, we present a brief discussion on the TMD evolution formalism for both the unpolarized and the polarized TMDs. 
 In Sec.~\ref{sec,asymmetry} we present  the $\sin(2\phi-\phi_{S})$ asymmetry in the pion-proton Drell-Yan process by employing the pion Boer-Mulders function and proton transversity TMD as nonperturbative inputs at the COMPASS kinematics. 
We provide a brief summary and conclusions in Sec.~\ref{concl}.

	\section{Pion TMDs}\label{pionTMDs}
	
	For a hadron, the quark TMDs are parameterized through the quark-quark correlation function~\cite{Meissner:2008ay,Pasquini:2014ppa,Bacchetta:2006tn} as,
	\begin{align}\label{coor}
		&\Phi_q^{[\Gamma]}(x,\vec{k}_\perp)=\frac{1}{2}\int\frac{\mathrm{d}z^-\mathrm{d}^2\vec{z}_\perp}{2(2\pi)^3}e^{ik\cdot z}\nonumber\\
		&\times\langle P,\,S|\bar\psi(0)\Gamma\mathcal{W}(0,z)\psi(z)|P,\,S\rangle|_{z^+=0}
	\end{align}
where $k^+=xP^+$ and $\vec{k}_\perp$ are the longitudinal and the transverse momenta of the struck quark, respectively. $|P,\,S\rangle$ is the bound state of the target hadron with mass $M$, momenta $(P^+,\vec{P}_\perp)$, where the transverse momentum $\vec{P}_\perp=\vec{0}$~\cite{Collins:1992kk}, and spin $S$. The Dirac matrix $\Gamma$ governs the Lorentz structure of the correlator $\Phi_q^{[\Gamma]}$ and its `twist' $\tau$ \cite{Jaffe:1991kp}. The Wilson line $\mathcal{W}$ maintains the color gauge invariance of the bilocal quark field operators in the correlation function~\cite{Bacchetta:2020vty}. 

For the pion, there are two leading twist TMDs namely the unpolarized quark TMD, $f^q_{1,\pi}(x,k_\perp^2)$, and the polarized quark TMD, $h_{1,\pi}^{\perp q}(x,k_\perp^2)$, also known as the pion Boer-Mulders function. They are defined through the parameterizations of the quark-quark correlator with $\Gamma\equiv\gamma^{+},\,\sigma^{i+} \gamma_{5}$, respectively,
\begin{align}
  \frac{1}{2}\text{Tr}[\Phi^{[\gamma^{+}]}]&=f^q_{1,\pi}(x,k_\perp^2), \label{tw-2TMDs1}\\
  \frac{1}{2}\text{Tr}[\Phi^{[i \sigma^{i+} \gamma_{5}]}]&=\frac{ \varepsilon_{T}^{i j} k_{\perp}^j}{M_{\pi}} {h_{1,\pi}^{\perp q}}(x,k_\perp^2)\label{tw-2TMDs2},
\end{align}
where $\epsilon_T^{11}=\epsilon_T^{22}=0$, and $\epsilon_T^{12}=-\epsilon_{T}^{21}=1$.

Ignoring the gauge link, we obtain the explicit expressions of the unpolarized pion TMD $f_{1,\pi}^{q}(x,k_\perp^2)$ as
\begin{equation}
 	f_{1,\pi}^{q}(x,  k_\perp^2)=\frac{1}{16\pi^3} \sum_{h,\bar{h}} |\Psi_{h \bar{h}}(x,\vec{k}_\perp)|^2  \;,
 \label{hf1}
 \end{equation} 
 where $\Psi_{h \bar{h}}(x,\vec{k}_\perp)$ is the pion light-front wavefunction in the momentum space with $h\, (\bar{h})$ being the helicity of the quark (antiquark) in the leading Fock sector. The pion unpolarized TMD, $f_{1,\pi}^{q}(x,k_\perp^2)$ satisfies the following normalization condition,
 \begin{equation}
 	\int \mathrm{d} x \,\mathrm{d}^2 \vec{k}_\perp \,f_{1,\pi}^{q}(x,k_\perp^2)=1  \;.
 \label{PDF-norm}
 \end{equation}
 
Meanwhile, to produce the nonzero pion Boer-Mulders function $h_{1,\pi}^{\perp q}(x,{k}_\perp^2)$, we need to consider the gauge link, which is physically equivalent to taking into account the initial (final)-state interactions of the active parton with the target remnant. We refer this collectively as gluon rescattering karnel. The pion  Boer-Mulders function is then expressed as~\cite{Ahmady:2019yvo},
\begin{align}
&	k_\perp^2 h_{1,\pi}^{\perp q}(x, k_\perp^2) = M_\pi \int \frac{\mathrm{d}^2 \vec{k}_\perp^{\prime}}{16\pi^3}~ i G(x, \vec{k}_\perp-\vec{k}_\perp^{\prime}) \nonumber\\
	&\times \sum_{h,\bar{h}} \Psi_{-h,\bar{h}}^*(x, \vec{k}^{\prime}_\perp) h k_\perp e^{i h \theta_{k_\perp}} \Psi_{h,\bar{h}}(x,\vec{k}_\perp) \;,
\label{BM-overlap}
\end{align}
where $k_{\perp}=|\vec{k}_{\perp}|$ and $G(x, \,\vec{k}_\perp-\vec{k}_\perp^{\prime})$ represents the gluon rescattering kernel with $(\vec{k}_\perp-\vec{k}_\perp^{\prime})$ being the transverse momentum carried by the exchanged gluon. The simplest approach is to assume that the perturbative Abelian gluon rescattering kernel, which is given by~\cite{Bacchetta:2008af,Wang:2017onm}
\begin{equation}
 	i G^{\mathrm{pert.}}(x, {\vec{k}_\perp-\vec{k}_\perp^{\prime}}) = \frac{C_F\alpha_s}{2\pi}\frac{1}{(\vec{k}_\perp-\vec{k}_\perp^{\prime})^2} \;,
 	\label{pert-G}
\end{equation}
with $\alpha_s$ being the fixed coupling constant and  the color factor $C_F=4/3$. 
An exact computation of nonperturbative gluon rescattering kernel is yet not available and, in practice, some approximation scheme is required.  Meanwhile, in terms of the so-called QCD lensing function $I(x, \vec{k}_\perp-\vec{k}_\perp^{\prime})$, the gluon rescattering kernel can be expressed as~\cite{Ahmady:2019yvo},
\begin{equation}
i G(x, \vec{k}_\perp-\vec{k}_\perp^{\prime})= -\frac{2}{(2\pi)^2} \frac{(1-x) I(x, \vec{k}_\perp-\vec{k}_\perp^{\prime})}{(\vec{k}_\perp-\vec{k}_\perp^{\prime})} \;,
	\label{Relation-GI}
\end{equation}
which has been derived from the relation between the chiral-odd GPD and the first moment of the pion  Boer-Mulders function~\cite{Burkardt:2007xm}.
In Ref. \cite{Gamberg:2009uk}, Gamberg and Schlegel derived the QCD lensing function \cite{Burkardt:2007xm} from the eikonal amplitude for final-state
rescattering via the exchange of non-Abelian SU$(3)$ soft gluons. The nonperturbative gluon rescattering kernel  derived in Ref.~\cite{Ahmady:2019yvo} has been successfully employed to compute T-odd TMDs of the pion~\cite{Ahmady:2019yvo,Kou:2023ady} and the proton~\cite{Gurjar:2022rcl}.

To compute the pion's leading twist TMDs, we employ the spin-improved holographic wave function, which is given by~\cite{Ahmady:2018muv,Ahmady:2019yvo} 
	\begin{align}
	 	\Psi_{h,\bar{h}}(x,\vec{k}_\perp)=& \Big[ (M_{\pi} x(1-x) + B m_q) h\delta_{h,-\bar{h}}  \nonumber\\
	 	&- B    k_\perp e^{-ih\theta_{k_\perp}}\delta_{h,\bar{h}}	\Big] \frac{\Psi (x, \vec{k}_\perp)}{x(1-x)}\,.
	 \label{spin-improved-wfn-k}
	 \end{align}
	 with 
	 	 \begin{equation}
	 	\Psi (x,\vec{k}_\perp)=\mathcal{N} \frac{1}{\sqrt{x (1-x)}}  \exp{ \Big[ -\frac{k_\perp^2+m_q^2}{2\kappa^2 x(1-x)} \Big] } \,,
 \label{hWF-k}
 \end{equation}
 where $m_q$ is the quark mass and $\mathcal{N}$ is a normalization constant fixed using
 \begin{equation}
 	\sum_{h,\bar{h}}\int \mathrm{d} x \frac{\mathrm{d}^2 \vec{k}_\perp}{16\pi^3} |\Psi_{h \bar{h}}(x,\vec{k}_\perp)|^2 =1 \;.
\end{equation}  
The parameter $B$ in Eq.~\eqref{spin-improved-wfn-k} is referred as the dynamical spin parameter. $B \to 0$ implies no spin-orbit correlations as in the original holographic wave function~\cite{Brodsky:2014yha,Brodsky:2008pf}, while, on the other hand, $B \ge 1$ represents a maximal spin-orbit correlation. With $B \ge 1$, $m_{q}=330$ MeV and $\kappa=523$ MeV, the pion wave function has been successfully employed to compute a wide
class of different and related pion observables, e.g., the electromagnetic form factors and associated radii, transition form factor, parton
distribution function (PDF), TMDs, etc., with remarkable
overall success~\cite{Ahmady:2016ufq,Ahmady:2018muv,Ahmady:2019yvo}.

	\begin{figure}
		\includegraphics[scale=0.55]{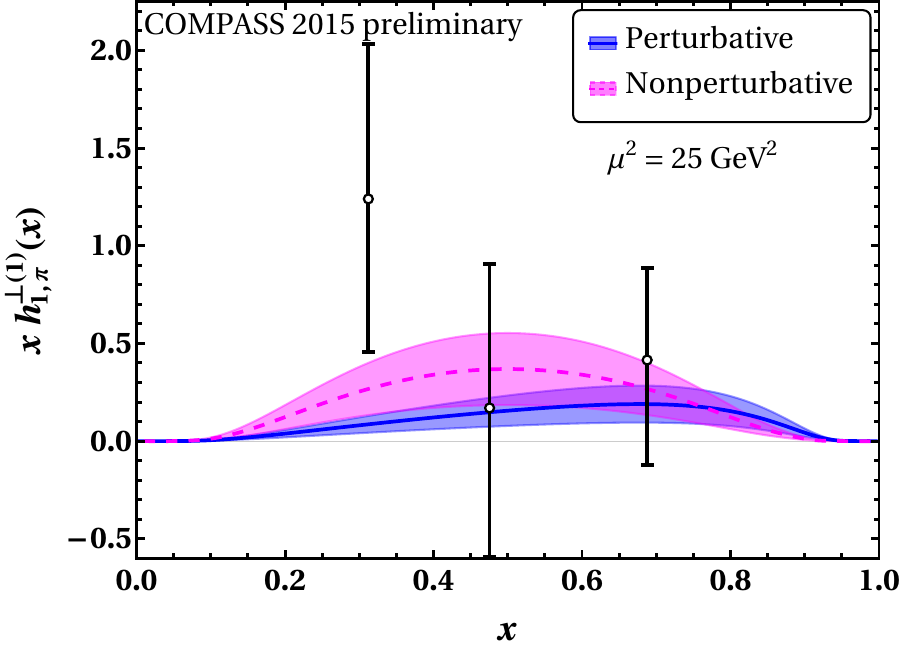}
		\caption{The $x$ dependence of the first moment of pion Boer-Mulders TMD at the scale $\mu^{2} = 25\rm{GeV}^{2}$. The blue and magenta lines correspond to the results generated by using the perturbative and nonperturbative gluon rescattering karnels, respectively. The uncertainty bands in our results are due to the uncertainties in the model parameters, $\kappa = 523 \pm 24$ MeV and $m_{f} = 330 \pm 16$ MeV. Our predictions are compared with the available COMPASS 2015 preliminary data~\cite{Longo:2019bih}.}
		\label{fig:BMTMDmoments}
	\end{figure}

With the pion wave function given in Eq.~\eqref{spin-improved-wfn-k}, the
explicit expression for the unpolarized quark TMD reads~\cite{Ahmady:2019yvo}
	\begin{align}\label{eq:pionunpolTMD}
		f_{1,\pi}^{q}(x,\kp^2)=&\frac{2 \mathcal{N}^{2}}{16\pi^{3}}\frac{(M_{\pi}x(1-x)+Bm_{q})^{2}+B^{2}\kp^{2}}{(x(1-x))^{3}}\nonumber\\&\times \exp\Big[-\frac{\kp^{2}+m_{q}^{2}}{x(1-x)\kappa^{2}}\Big]\,.
  \end{align}
If we employ the perturbative gluon rescattering kernel, Eq.~\eqref{pert-G}, we obtain an analytical expression for the pion Boer-Mulders function as,
	\begin{align}\label{eq:pionBMTMD}
&		h_{1,\pi}^{\perp q}(x,\kp^{2})=\alpha_{s}B C_{F}\frac{M_{\pi}\mathcal{N}^{2}}{4\pi^{3}}\frac{M_{\pi}x(1-x)+Bm_{q}}{(x(1-x))^{2}}\Big(\frac{\kappa}{\kp}\Big)^{2}\nonumber \\&\times\exp\Big[-\frac{\kp^{2}+2m_{q}^{2}}{2\kappa^{2}x(1-x)}\Big]\Big(1-\exp\Big[-\frac{\kp^{2}}{2\kappa^{2}x(1-x)}\Big]\Big)\,.
	\end{align}
Note that if $B \to 0$, the holographic Boer-Mulders function vanishes. Meanwhile, it is hardly sensitive to the value of $B$  for $B \ge 1$, since the wave function normalization constant $\mathcal{N} \sim 1/B^2$ when $B \ge 1$. Using a nonperturbative gluon rescattering kernels~\cite{Ahmady:2019yvo}, the pion Boer-Mulders can not be expressed analytically. We then compute it numerically. 

  In Fig. \ref{fig:BMTMDmoments}, we illustrate the differences between the first moment of the holographic pion Boer-Mulders function:
  \begin{align}
  h_{1,\pi}^{\perp (1)}(x)=\int \mathrm{d}^2 \vec{k}_\perp \, \frac{\vec{k}_\perp^2}{2M_{\pi}^2}\, h_{1,\pi}^{\perp q}(x,\kp^{2})\,,
  \end{align}
  generated by the perturbative and nonperturbative kernels. The perturbative result is obtained with the coupling constant $\alpha_s=0.3$. We evolve our pion TMDs using the Collins-Soper TMD evolution prescription~\cite{Collins:2011zzd,Boussarie:2023izj} 
  (see Sec.~\ref{TMDevolsec}) and compare our results for $h_{1,\pi}^{\perp (1)}(x)$ with the available COMPASS 2015 preliminary experimental data at $\mu^{2} = 25$ GeV$^{2}$~\cite{Longo:2019bih}. We find an acceptable compatibility for both the perturbatively and nonperturbatively generated results with the COMPASS data considering their large uncertainties. However, it becomes apparent that the nonperturbative kernel does a better job, bringing our predictions closer to the  experimental data.

	\section{Proton TMDs}\label{ProtonTMDs}
In this section, we briefly discuss about the leading twist T-even TMDs of the proton in a light-front quark-diquark model, where the proton wavefunctions are constructed from the solution of soft-wall anti–de Sitter (AdS)/QCD~\cite{Maji:2016yqo}. 
In this model, the proton state is expressed as a two-particle bound state of a quark 
and a diquark within a spin-flavor SU$(4)$ structure:
 \begin{align}
 	|P; \pm\rangle = C_S|u~ S^0\rangle^\pm + C_V|u~ A^0\rangle^\pm + C_{VV}|d~ A^1\rangle^\pm. \label{PS_state}
 \end{align} 
where $\mid u~ S^0\rangle$, $|u~ A^0\rangle$, and $|d~ A^1\rangle$ are two particle states with isoscalar-scalar, isoscalar-axialvector 
and isovector-axialvector diquark, respectively~\cite{Jakob:1997wg,Bacchetta:2008af}. The proton states with
helicities plus and minus correspond to the states with $J^z=+\frac{1}{2}$ and $J^z=-\frac{1}{2}$ , respectively. The modified form of the soft-wall AdS/QCD wave functions for two particle Fock state is given by~\cite{Gutsche:2013zia,Mondal:2015uha,Maji:2016yqo}
\begin{align}
\varphi_i^{(\nu)}(x,\vec{k}_\perp)=&\frac{4\pi}{\kappa}\sqrt{\frac{\log(1/x)}{1-x}}x^{a_i^\nu}(1-x)^{b_i^\nu}\nonumber\\
&\times\exp\Big[-\delta^\nu\frac{\vec{k}_\perp^2}{2\kappa^2}\frac{\log(1/x)}{(1-x)^2}\Big].
\label{LFWF_phi}
\end{align}
here $\nu$ stands for the quark flavors inside the proton. The explicit form of the wave functions in the AdS/QCD inspired quark-diquark model is given in Ref.~\cite{Maji:2016yqo}. The AdS/QCD scale parameter is taken as $\kappa =0.4$ GeV as determined in Ref.~\cite{Chakrabarti:2013gra}. The parameters $a_i^\nu,b_i^\nu$ and $\delta^\nu$ are fixed by fitting the nucleon form factors~\cite{Maji:2016yqo}. 

At the leading twist, the TMD correlator, Eq.~\eqref{coor}, for the proton is connected with the corresponding eight TMDs for different Dirac structures as~\cite{Boer:1997nt,Bacchetta:2006tn,Meissner:2009ww}
\begin{align}
    \Phi^{\left[\gamma^{+}\right]}&\left(x, \vec k_{\perp} \right)=f_{1}-\frac{\epsilon_{\perp}^{i j} k_\perp^{i} S_\perp^{j}}{M} f_{1 T}^{\perp}, \label{TMDs1}\\
    \Phi^{\left[\gamma^{+} \gamma^{5}\right]}&\left(x, \vec k_{\perp}\right)=S^{3} g_{1 L}+\frac{\vec k_{\perp} \cdot \vec S_\perp}{M} g_{1 T}, \label{TMDs2}\\
    \Phi^{\left[i \sigma^{j+} \gamma^{5}\right]}&\left(x, \vec k_{\perp}\right)=S_\perp^{j} h_{1}+S^{3} \frac{k_\perp^{j}}{M} h_{1 L}^{\perp}\nonumber\\
    &+S_\perp^{i} \frac{2 k_\perp^{i} k_\perp^{j}-\left(\vec k_{\perp}\right)^{2} \delta^{i j}}{2 M^{2}} h_{1 T}^{\perp}+\frac{\epsilon_{\perp}^{j i} k_\perp^{i}}{M} h_{1}^{\perp},
    \label{TMDs3}
\end{align}
where $i,j=1,2$ and antisymmetric tensor $\epsilon^{12}_\perp=-\epsilon^{21}_\perp=1$. $S^3$ and $S_\perp$ correspond to the helicity and transverse component of the proton's spin, respectively.

 \begin{figure}
	\includegraphics[scale=0.55]{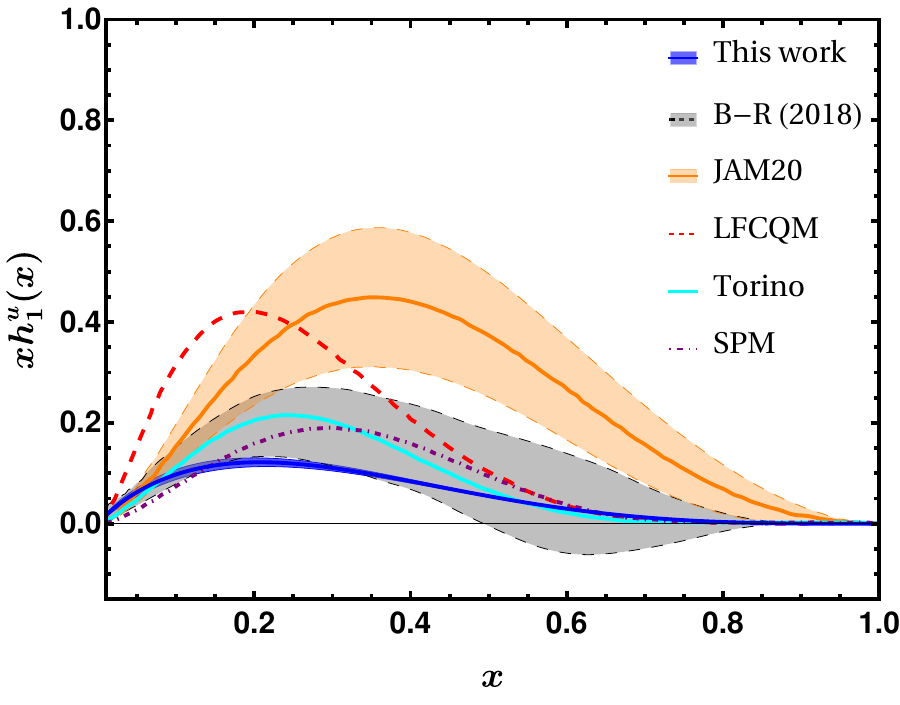}\vspace{0.2cm}
		\includegraphics[scale=0.55]{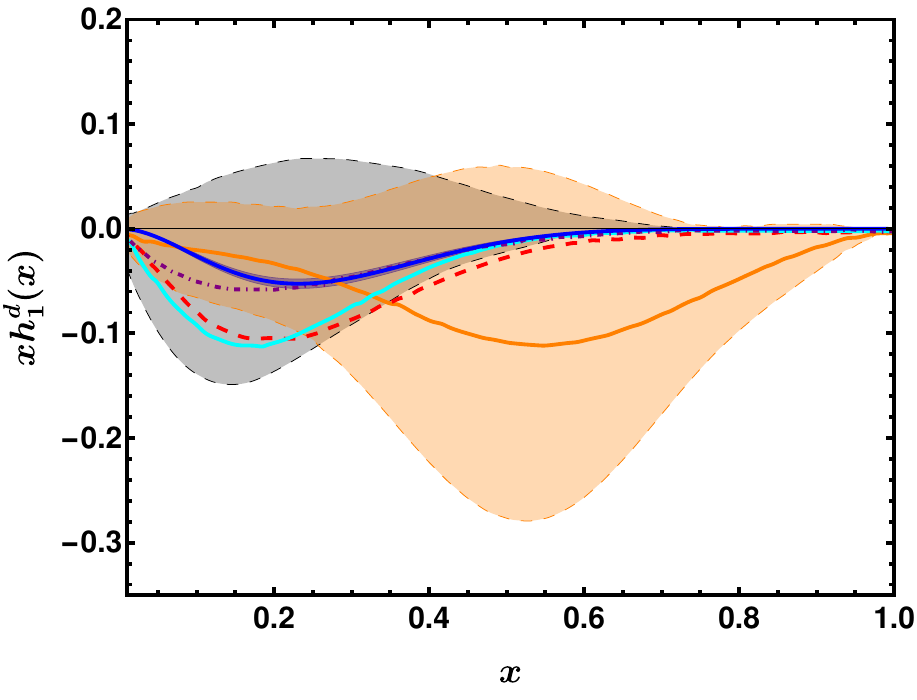}
		\caption{The transversity distribution $xh_{1}^{\nu}(x)$ as a function of $x$ at $Q^{2}=2.4$ GeV$^{2}$ for the up (upper panel) and down (lower panel) quarks. Our results (blue bands) are compared with various global analyses and model computations. The gray-bands with dashed borders correspond to the global fits from Bacchetta-Radici (2018)~\cite{Radici:2018iag}. The orange bands with orange solid lines represent the global analyses from JAM20 Collaboration~\cite{Cammarota:2020qcw}, while red-dashed, solid-cyan and purple-dashed curves correspond to the light-front constitute quark model (LFCQM)~\cite{Pasquini:2008ax,Boffi:2009sh,Pasquini:2011tk}, Torino extractions~\cite{Anselmino:2013vqa} and spectator model (SPM)~\cite{Gamberg:2007wm}, respectively. }  
		\label{fig:quarkTransversity}
	\end{figure}

Using the light-front wave functions of the quark-diquark model inspired by soft-wall AdS/QCD~\cite{Maji:2016yqo} in the correlator of Eq.~\eqref{coor} and comparing with the parameterizations in Eqs.~\eqref{TMDs1}-\eqref{TMDs3}, the unpolarized and the transversity TMDs contributing to the $\sin(2\phi-\phi_{s})$ azimuthal asymmetry reads explicitly as
	\begin{align}\label{unpolTMD}
		&f_{1,p}^{\nu}(x,\kp^{2})=\Big(C_{S}^{2}N_{S}^{\nu 2}+C_{V}^{2}\big(\frac{1}{3}N_{0}^{\nu 2}+\frac{2}{3}N_{1}^{\nu 2}\big)\Big)\frac{\ln(1/x)}{\pi\kappa^{2}}\nonumber\\&\times\Big[T_{1}^{\nu}(x)+\frac{\kp^{2}}{M^{2}}T_{2}^{\nu}(x)\Big]\exp\Big[-R^{\nu}(x)\kp^{2}\Big]\,,
		\end{align}
		and
		\begin{align}
\label{protontransversity}
	&	h_{1,p}^{\nu}(x,\kp^{2})=\Big(C_{S}^{2}N_{S}^{\nu 2}-C_{V}^{2}\frac{1}{3}N_{0}^{\nu 2}\Big)\frac{\ln(1/x)}{\pi\kappa^{2}}\nonumber\\&\times T_{1}^{\nu}(x)\exp\Big[-R^{\nu}(x)\kp^{2}\Big]\,,
	\end{align}
respectively,	where $T_{1}^{\nu}(x)$, $T_{2}^{\nu}(x)$ and $R^{\nu}(x)$ are given by,
	\begin{eqnarray}
		T_{1}^{\nu}(x)&=&x^{2a_{1}^{\nu}}(1-x)^{2b_{1}^{\nu}-1}\,,\nonumber\\
		T_{2}^{\nu}(x)&=&x^{2a_{2}^{\nu}-2}(1-x)^{2b_{2}^{\nu}-1}\,,\nonumber\\
		R^{\nu}(x)&=&\delta^{\nu}\frac{\ln(1/x)}{\kappa^{2}(1-x)^{2}}\,.
	\end{eqnarray}
All the model parameters can be found in Ref.~\cite{Maji:2016yqo}. The prefactors containing $C_j\,(j=S,V,VV)$ together with the normalization constants $N_r(r=S,0,1)$ satisfy the quark counting rules for unpolarized TMDs.

	In Fig.~\ref{fig:quarkTransversity} we show the quark transversity distribution for the up (upper panel) and the down (lower panel) quarks, respectively. We compared the quark-diquark model transversity PDFs at the scale $\mu^{2}= 2.4$ GeV$^{2}$ with the recently extracted fits from the global analysis  by Bacchetta and Radici~\cite{Radici:2018iag}. In addition, we perform a comparative analysis with the JAM20 global fits~\cite{Cammarota:2020qcw}, the results from the LFCQM model~\cite{Pasquini:2008ax,Boffi:2009sh,Pasquini:2011tk} and the SPM model~\cite{Gamberg:2007wm} as well as with the Torino extractions~\cite{Anselmino:2013vqa}. 
	We observe that the quark transversity distributions in the quark-diquark model are more or less consistent with the Bacchetta-Radici global fits~\cite{Radici:2018iag}, however, they are somewhat underestimated compared to the other global analyses and model predictions.

\section{Evolution of TMDs}\label{TMDevolsec}
In order to compare the model calculations of asymmetry with experimental data, it is necessary to evolve the distribution functions. In this section, we focus on the evolution formalism of TMDs by using the TMD factorization process~\cite{Collins:1984kg,Collins:2011zzd}. The scale evolution of TMDs can be done conveniently in coordinate ($b_\perp$) space. The distribution functions in $b_\perp$-space are obtained by performing  Fourier transformation of the TMDs with respect to the parton transverse momentum $\kp$~\cite{Aybat:2011zv},
\begin{align}
	\widetilde{F}(x,b_\perp)=\int_{0}^{\infty}d{k}_{\perp}{k}_{\perp}  J_{0}({k}_{\perp} \bp)F(x,{k}_{\perp}).
\end{align}
where $\widetilde{F}(x,b_{\perp})$ and $F(x,k_{\perp})$ are the distribution functions in the position as well as momentum spaces, respectively. The TMD evolution for the unpolarized distributions of the proton and the pion in $b_{\perp}$-space can be obtained by solving the Collins-Soper (CS) and renormalization
group (RG) evolution equations and the general solution for the energy dependence of $\widetilde{f}_{1,q}(x,b_{\perp})$ is given by~\cite{Echevarria:2012pw,Echevarria:2014xaa,Echevarria:2014rua},
\begin{align}
	&\widetilde{f}_{1,q}(x,\bp;Q_{f})\nonumber\\
	& =\widetilde{f}_{1,q}(x,\bp)R_{\mathrm{pert}}(Q_{f},Q_{i},b_{\ast})R_{\mathrm{NP}}(Q_{f},\bp),
	\label{eq:f_b}
\end{align}
where 
$\widetilde{f}_{1,q}(x,\bp)$ is the unpolarized TMD  in the $\bp$-space at the model scale,
$R_{\mathrm{pert}}(Q_{f},Q_{i},b_{\ast})$ and $R_{\mathrm{NP}}(Q_{f},\bp)$ are the perturbative and nonperturbative evolution kernels of TMDs, respectively. $Q_{i}=2e^{-\gamma_{E}}/b_{\ast}$ (with $\gamma_{E}\simeq 0.577$) with 
 the  choice of $b_{\ast}$ in such a way that $b_{\ast}(\bp)=\bp/({1+\frac{\bp^{2}}{b_{\textrm{max}}^{2}}})^{1/2} \simeq{b_{\textrm{max}}}$  at $\bp\rightarrow\infty$ and $b_{\ast}(\bp)\simeq \bp$ when $\bp \rightarrow 0$~\cite{Collins:2014jpa}. These allow one to avoid hitting the Landau pole by freezing the scale $\bp$~\cite{Collins:1984kg}. $b_{\text{max}}$ seprates the perturbative and nonperturbative regions of the TMDs and it is fixed phenomenologically as $b_{\textrm{max}}=1.5$ GeV$^{-1}$~\cite{Sun:2014dqm}. The perturbative part of the evolution kernel $R_{\mathrm{\textrm{pert}}}(Q_{f},Q_{i},b_{\ast})$ is same for all type of distributions, i.e., it is spin independent and  has the following form~\cite{Echevarria:2014xaa,Kang:2011mr,Aybat:2011ge,Echevarria:2012pw,Echevarria:2014rua}:
\begin{align}
	\label{eq:Rperturbative}
	&R_{\mathrm{\textrm{pert}}}(Q_{f},Q_{i},b_{\ast})\\
	&=\exp\left\{-\int^{Q_{f}}_{Q_i}\frac{d\bar{\mu}}{\bar{\mu}}\left[A\left(\alpha_s(\bar{\mu})\right)
	\mathrm{ln}\frac{Q_{f}^2}{\bar{\mu}^2}+B(\alpha_s(\bar{\mu}))\right]\right\},\nonumber
\end{align}
where the coefficients $A$ and $B$ in Eq.~(\ref{eq:Rperturbative}) can be expanded perturbatively as,
\begin{align}
	A=\sum_{n=1}^{\infty}A^{(n)}(\frac{\alpha_s}{\pi})^n, ~~~~
	B=\sum_{n=1}^{\infty}B^{(n)}(\frac{\alpha_s}{\pi})^n
\end{align}
with coefficients $A^{(n)}$ and $B^{(n)}$ corresponding to next-to-leading-logarithmic (NLL) accuracy~\cite{Collins:1984kg,Aybat:2011zv,Echevarria:2012pw},
\begin{align}
	A^{(1)}&=C_F,\\
	A^{(2)}&=\frac{C_F}{2}\left[C_A\left(\frac{67}{18}-\frac{\pi^2}{6}\right)-\frac{10}{9}T_Rn_f\right],\\
	B^{(1)}&=-\frac{3}{2}C_F.
\end{align} 
Here $C_{F}$ and $n_{f}$ are the color factor and number of flavors in the hadron, respectively.

Meanwhile, the nonperturbative part of the evolution kernel in Eq.~(\ref{eq:f_b}) has been studied phenomenologically. In Ref.~\cite{Collins:1984kg}, a generic form for the nonperturbative evolution kernel, $R_{\rm{NP}}(Q_{f};\bp)$, was proposed,
\begin{equation}
	\label{eq:snp_gene}
	R_{\rm{NP}}(Q_{f};\bp)=\exp\left\{-\left[g_1(\bp)+g_2(\bp)\ln \frac{Q_{f}}{Q_0}\right] \right\},
\end{equation}
where $Q_{0}$ is the model scale, $g_{1}(\bp)$ and $g_{2}(\bp)$ depend on the hadronic distribution functions and for the proton they are given by,
\begin{eqnarray}
	g_{1}^{p}(\bp)=\frac{g_{1}^{p}}{2}\bp^{2},~~~g_{2}^{p}(\bp)=\frac{g_{2}^{p}}{2}\ln\frac{\bp}{b_{\ast}}
\end{eqnarray}
with $g_{1}^{p}=0.212 \pm 0.006$ GeV$^2$ and $g_{2}^{p}= 0.84\pm 0.037$ GeV$^2$~\cite{CDF:2012brb,D0:2007lmg}. For the pion
\begin{eqnarray}
	g_{1}^{\pi}(\bp)={g_{1}^{\pi}}\bp^{2},~~~g_{2}^{\pi}(\bp)=g_{2}^{\pi}\ln\frac{\bp}{b_{\ast}}.
\end{eqnarray}
The numerical values of $g_1^\pi$ and $g_2^\pi$ are obtained by fitting to the $\pi^- N$ Drell-Yan data~\cite{Conway:1989fs}: 
 $g^{\pi}_{1}=0.082 \pm 0.022$ GeV$^2$ and $g^{\pi}_{2}=0.394 \pm 0.103$ GeV$^2$.
 After performing the scale evolution of the unpolarized TMD distributions in $\bp$-space, one  obtains the evolved distributions in $\kp$-space by taking the inverse Fourier transformation of $\widetilde{F}$ as,
\begin{align}
	f_{1,q}(x,\kp;Q_{f})=\int_{0}^{\infty}\frac{d^{2}\bp}{(2\pi)^{2}}J_{0}(\kp\bp)\widetilde{f}_{1,q}(x,\bp;Q_{f}).
\end{align}

Till date, a definitive method for the evolution of proton transversity TMD has not been established. However, in Ref.~\cite{Kang:2015msa,Bacchetta:2013pqa}, the authors have made notable progress by revealing that the scale evolution of the proton transversity TMD can be approached similarly to the evolution of the unpolarized distribution function as,
\begin{align}
	&\widetilde{h}_{1}(x,b_{\perp};Q_{f})=\widetilde{h}_{1}(x,\bp)R_{\mathrm{pert}}(Q_{f},Q_{i},b_{\ast})R_{\mathrm{NP}}(Q_{f},\bp),
	\label{eq:h_proton_b}
\end{align}
where $\widetilde{h}_{1}(x,\bp)$ is the transversity distribution in the coordinate space. The nonperturbative evolution kernel associated with the proton transversity distribution is also assumed to be the same as that for the unpolarized distribution function~\cite{Kang:2015msa}.\\

Finally, we look into the evolution of the pion Boer-Mulders distribution function. The Boer-Mulders function in the $b_{\perp}$-space can be defined as~\cite{Wang:2017onm},
\begin{align}
	\widetilde{h}_{1,q/\pi}^{\perp\alpha}(x,b_{\perp})=\int d^2\bm{k}_\perp e^{-i\bm{k}_\perp\cdot\bm{b}_\perp}\frac{k^\alpha_\perp}{M_\pi}
	h^{\perp q}_{1,\pi}(x,\bm{k}_\perp^{2}). \label{eq:pibm}
\end{align}
In the small $\bp$ region, the pion Boer-Mulders function $h_{1,\pi}^{\perp q}$ can also be written in terms of collinear chiral-odd twist-3 quark-gluon-quark correlation function $T^{(\sigma)}_{q/\pi,F}(x,x)$~\cite{Wang:2018naw} as,
\begin{align}
	\widetilde{h}_{1,q/\pi}^{\perp\alpha}(x,b_{\perp})=(\frac{-ib_\perp^\alpha}{2})T^{(\sigma)}_{q/\pi,F}(x,x).
\end{align}
The collinear distribution $T^{(\sigma)}_{q/\pi,F}(x,x)$, known as Qiu-Sterman function, is related to the first transverse moment of the Boer-Mulders function $h_{1,q/\pi}^{\perp (1)}$~\cite{Li:2019uhj,Anselmino:2012aa}, which is expressed as,
\begin{align}
	T^{(\sigma)}_{q/\pi,F}(x,x)&=\int d^2 \bm{k}_\perp\frac{\bm{k}_\perp^2}{M_\pi}h_{1,\pi}^{\perp q}(x,\bm{k}_\perp^{2})\nonumber\\
&	= 2M_\pi h_{1,q/\pi}^{\perp (1)}(x). \label{eq:qsbm}
\end{align}
The nonperturbative evolution kernel for the pion Boer-Mulders function is still unknown. Here, we assumed that it is same as the unpolarized one, 
i.e., $R_\mathrm{NP}^{h_{1,\pi}^{\perp q}}=R_\mathrm{NP}^{f_{1,\pi}^{q}}$ as mentioned in Ref.~\cite{Li:2019uhj}.
Therefore, we can obtain the evolved Boer-Mulders function of the pion in $b_{\perp}$-space as,
\begin{align}
	\widetilde{h}_{1,q/\pi}^{\perp \alpha}(x,b_{\perp};Q_{f})&=-\frac{ib_\perp^\alpha}{2}T^{(\sigma)}_{q/\pi,F}(x,x)
	\nonumber \\&
	\times R_{\mathrm{pert}}(Q_{f},Q_{i},b_{\ast})R_{\mathrm{NP}}(Q_{f},\bp).
	\label{eq:BM_b}
\end{align}
The pion Boer-Mulders function can also be transformed into $\kp$ space by performing the inverse fourier transformation of $\widetilde{h}_{1,q/\pi}^{\perp \alpha}(x,b_{\perp};Q_{f})$ as,
\begin{align}
	\frac{\kp}{M_{\pi}}&h_{1\pi}^{\perp q}(x,\kp;Q_{f})=\int_{0}^{\infty}\bp\frac{d^{2}\bp}{(2\pi)^{2}}J_{1}(\kp\bp)
\nonumber \\& 
~~\times R_{\mathrm{pert}}(Q_{f},Q_{i},b_{\ast})
	 R_{\mathrm{NP}}(Q_{f},\bp)	
	h_{1,q/\pi}^{\perp(1)}(x).
\end{align}

	\section{$\sin(2\phi-\phi_{s})$ azimuthal asymmetry}
\label{sec,asymmetry}
 The computation of the $\sin(2\phi-\phi_{s})$ asymmetry results from the convolution of the Boer-Mulders function of the pion beam and the transversity distribution of the proton target at leading twist and taking into account the scale evolution effects of the TMDs. 
The process we investigate,	the pion-induced Drell-Yan process, is expressed as,
	\begin{eqnarray}\label{DYprocess}\nonumber
		\pi^{-}(P_{\pi})+p^{\uparrow}(P_{p}) &\rightarrow& \gamma^{\star}(q)+X\\
		&\rightarrow& l^{+}(\ell)+l^{-}(\ell^{\prime})+X\,,
	\end{eqnarray}
	where $P_{\pi}$, $P_{p}$, and $q$ stand for the incoming four-momenta of the pion, the target proton, and the virtual photon, respectively. 
	The uparrow ($\uparrow$) stands for the transverse polarization of the target. The experimental observables are characterized by the following kinematic variables: 
	\begin{align}\label{kinematics}
		&s=(P_{\pi}+P_{p})^{2},\quad x_{\pi}=\frac{Q^{2}}{2P_{\pi}.q},\quad x_{p}=\frac{Q^{2}}{2P_{p}.q},
		\nonumber \\
  &x_{F}=x_{\pi}-x_{p}=2q_{L}/s,\quad \tau=Q^{2}/s=x_{\pi}x_{p},\nonumber \\
  & y=\frac{1}{2}\ln\frac{q^{+}}{q^{-}}=\frac{1}{2}\ln\frac{x_{\pi}}{x_{p}}\,,
	\end{align}
	where $s$ is square of the total center-of-mass energy, $x_{\pi}$ and $x_{p}$ are the Bjorken variables of the incoming pion and the target proton, respectively. The longitudinal momentum of the virtual photon in the incident hadron c.m. frame is denoted by $q_L$.  $x_F$ is the Feynman variable and $y$ is the lepton pair rapidity. 
	
	The differential cross section in $\pi$-p Drell-Yan process for a transversely polarized target is described by the following generic form~\cite{COMPASS:2010shj,Arnold:2008kf},
	\begin{align}\label{differentialcorssection}
		&\frac{d\sigma}{d^{4}qd\Omega}= \frac{\alpha^{2}_{em}}{Fq^{2}}\hat{\sigma}_{U}\Bigl\{(1+D_{[\sin^{2}\theta]}A_{U}^{\cos 2\phi}\cos 2\phi)\nonumber\\
		&+|S_{T}|\Big[A_{T}^{\sin \phi_{s}}\sin \phi_{s}+
		D_{[\sin^{2} \theta]}\Big(A_{T}^{\sin(2\phi+\phi_{s})}\sin(2\phi+\phi_{s})\nonumber\\
 & +A_{T}^{\sin(2\phi-\phi_{s})}\sin(2\phi-\phi_{s})\Big)\Big]\Bigr\}\,.	
	\end{align} 
	In the above Eq.~\eqref{differentialcorssection}, the azimuthal angle of the target polarization vector $S_{T}$ in the target rest frame is denoted by $\phi_{s}$, and the azimuthal and polar angles of the lepton momentum in the Collins-Soper frame~\cite{Peng:2018tty} are denoted by $\phi$ and $\theta$, respectively. In the Collins-Soper frame, $\hat{\sigma}_{U}$ is given by  $\hat{\sigma}_{U}=F_{U}^{1}(1+\cos^{2}\theta)$ with $F_{U}^{1}$ being the unpolarized structure function. The $D_{[f(\theta)]}$ denotes the depolarization factor, which depends only on $\theta$, and at leading-order (LO), it is reduced to $\sin^{2}\theta/(1+\cos^{2}\theta)$. Furthermore, $A_{P}^{f[\phi,\phi_{s}]}$ stands for the azimuthal asymmetry with a $f[\phi,\phi_{s}]$ modulation, where $P=U$ or $T$ stands for the polarization of the target proton ($U$: unpolarized, $T$: transversely polarized).
	The ratio between the related structure function $F_{P}^{f[\phi,\phi_{s}]}$ and the unpolarized structure function $F_{U}^{1}$ can be used to express the asymmetry $A_{P}^{f[\phi,\phi_{s}]}$. Here, we emphasize on the $\sin(2\phi-\phi_{s})$ weighted asymmetry, which is defined as,
	\begin{eqnarray}
		A_{T}^{\sin (2\phi-\phi_s)}(x_p,x_\pi,q_{\perp})=\frac{F_{T}^{\sin (2\phi-\phi_s)}(x_p,x_\pi,q_{\perp})}{F_{U}^{1}(x_p,x_\pi,q_{\perp})}\,,
		\label{eq:asymmetry}
	\end{eqnarray}
	where the denominator, $F_{U}^{1}(x_p,x_\pi,q_{\perp})$, is the convolution of the unpolarized distribution functions from each hadron,
	\begin{eqnarray}
		F_{U}^{1}=\mathcal{C}\left[f_{1,\bar{q}/\pi} f_{1,q/p}\right],
		\label{eq:FUU}
	\end{eqnarray}
	and the numerator, $F_{T}^{\sin (2\phi-\phi_{s})}(x_p,x_\pi,q_{\perp})$, is the convolution of the pion Boer-Mulders TMD and the proton transversity distributions~\cite{Li:2019uhj,Bastami:2020asv},
	\begin{eqnarray}
		F_{T}^{\sin (2\phi-\phi_S)}&=-\mathcal{C}\left[\frac{\hat{h}\cdot\vec{k}_{\perp\pi}}{M_\pi}h_{1,\bar{q}/\pi}^\perp h_{1,{q}/p}\right],
		\label{eq:FUT}
	\end{eqnarray}
	with $\hat{h}=\vec{q}_\perp/{|\vec{q}_\perp|}$.
	The convolution of unpolarized TMDs in Eq.~(\ref{eq:FUU}) is defined in the momentum space as~\cite{Arnold:2008kf},
	\begin{align}
		&\mathcal{C}[\omega(\vec{k}_{\perp\pi},\vec{k}_{\perp p})f_{1,\bar{q}/\pi}f_{1,q/p}]=\frac{1}{N_{c}}\sum_{q}e_{q}^{2}\nonumber\\
&	\times	\int d^{2}\vec{k}_{\perp\pi}d^{2}\vec{k}_{\perp p}   \delta^{2}(\vec{k}_{\perp\pi}+\vec{k}_{\perp p}-\vec{q}_\perp)
		\omega(\vec{k}_{\perp\pi},\vec{k}_{\perp p}) \nonumber\\  &\times \Big[f_{1,\bar{q}/\pi}(x_\pi,{k}^2_{\perp\pi}) f_{1,q/p}(x_p,{k}^2_{\perp p})
		\Big],
		\label{eq:C}
	\end{align}
	where $e_{q}$ is the fractional charges of the flavors; $N_c=3$ is the number of colors; $\vec{q}_\perp$, $\vec{k}_{\perp\pi}$ and $\vec{k}_{\perp p}$ denote the transverse momenta of the lepton pair, antiquark and quark in the initial hadrons, respectively and $\omega(\vec{k}_{\perp\pi},\vec{k}_{\perp p})$ is weight factor, which projects out the corresponding azimuthal angular dependence. The sum over $q=u, \bar{u}, d$ and $\bar{d}$ includes the active flavors in the initial state hadrons. 
Using the property of the Fourier transformation,
	\begin{equation}
		\delta^{2}(\bm{k}_{\perp\pi}+\bm{k}_{\perp p}-\bm{q}_\perp) = \int \frac{d^2 \bm{b}_\perp}{(2\pi)^2} e^{- i \bm b_\perp\cdot(\bm{k}_{\perp\pi}+\bm{k}_{\perp p}-\bm{q}_\perp)},
	\end{equation}
	one can express explicitly the unpolarized structure function $F_{U}^{1}(x_{p},x_{\pi},q_{\perp})$ as~\cite{Collins:2014jpa},
	\begin{align}
		&F_{U}^{1}(x_{p},x_{\pi},q_{\perp})=\frac{1}{N_c}\sum_q e_q^2 \int d^{2}\bm{k}_{\perp\pi}d^{2}\bm{k}_{\perp p}\int\frac{d^2\bm b_\perp}{(2\pi)^2} \nonumber\\
        & \times e^{-i(\bm{k}_{\perp\pi}+\bm{k}_{\perp p}-\bm{q}_\perp)\cdot\bm{b_\perp}}f_{1,\bar{q}/\pi}(x_\pi,\bm{k}^{2}_{\perp\pi})f_{1,{q}/p}(x_p,\bm{k}^{2}_{\perp p})
		\nonumber\\
		&=\frac{1}{N_c}\sum_q e_q^2\int_0^\infty  \frac{b_{\perp} db_{\perp}}{2\pi}J_0(q_\perp b_{\perp})\widetilde{f}_{1,\bar{q}/\pi}(x_\pi,b_{\perp};Q_{f}) 
	\nonumber\\ &	\hspace{3.5cm}\times\widetilde{f}_{1,{q}/p}(x_p,b_{\perp};Q_{f}).
	\label{eq:final_FU}
	\end{align}
where $J_{0}$ is the Bessel function of zeroth order and $\widetilde{f}_{1,\bar{q}/\pi}$ and $\widetilde{f}_{1,{q}/p}$ are the unpolarized pion and proton evolved TMDs in position space, respectively.

Similarly, the spin dependent structure function $F_{T}^{\sin(2\phi-\phi_s)}(x_{p},x_{\pi},q_{\perp})$ can be written as, 
	\begin{align}
&		F_{T}^{\sin(2\phi-\phi_s)}(x_{p},x_{\pi},q_{\perp})
		= -\frac{1}{N_c}\sum_q e_q^2 \int d^{2}\bm{k}_{\perp\pi}d^{2}\bm{k}_{\perp p}\nonumber\\
		&
\times        \int\frac{d^2\bm{b}_\perp}{(2\pi)^2}
		e^{-i\bm{b}_\perp\cdot(\bm{k}_{\perp\pi}+\bm{k}_{\perp p}-\bm{q}_\perp)}
		\frac{\hat{h}\cdot\bm{k}_{\perp\pi}}{M_\pi}h_{1,\bar{q}/\pi}^\perp(x_\pi,\bm{k}^2_{\perp\pi})
		\nonumber\\&
      \hspace{5.4cm} \times 
		h_{1,{q}/p}(x_p,\bm{k}^2_{\perp p})
		\nonumber\\
		&=-\frac{1}{N_c}\sum_{q} e_q^2\int_0^\infty \frac{db_{\perp}}{{4\pi}}b_{\perp}^2J_1(q_\perp b_{\perp})
		\widetilde{h}_{1,q/p}(x_p,b_{\perp};Q_{f})
		\nonumber		 \\ &
	~~~~~~~\times	T^{(\sigma)}_{\bar{q}/\pi,F}(x,x)
 R_{\mathrm{pert}}(Q_{f},Q_{i},b_{\ast})R_{\mathrm{NP}}(Q_{f},\bp), 
		\label{eq:final FT}
	\end{align}
where $J_{1}$ is the modified Bessel function of first kind, $\widetilde{h}_{1,q/p}$ is the evolved proton transversity distribution in the position space, $T^{(\sigma)}_{\bar{q}/\pi,F}$ is the twist-3 Qiu-Sterman function and $R_{\text{pert}}$ and $R_{\text{NP}}$ are the perturbative and nonperturbative evolution kernels, respectively. 
The above structure functions also depend on the final evolution scale $Q_{f}$. Here, we do not indicate it in the expressions. Along with the evolution scale $Q_{f}$, 
the asymmetry and the structure functions, Eq.(\ref{eq:asymmetry}), are the functions of the variables $x_{\pi}, x_{p}, \text{and}~q_{\perp}$. However, while presenting the numerical results, we display the asymmetry as a function of one of the the above variables. It is obvious that the structure functions are integrated over the other variables within the accepted experimental domain. 
	
	We compute the $\sin(2\phi-\phi_{s})$ azimuthal asymmetry in the pion-induced transversely polarized Drell-Yan process at the kinematics of the COMPASS Drell-Yan program and
compare it with the recent experimental data~\cite{COMPASS:2017jbv}. We employ the distribution functions of the pion evaluated in Eqs.~(\ref{eq:pionunpolTMD}) and (\ref{eq:pionBMTMD}), as well as the distribution functions of the proton
	target given in Eqs.~(\ref{unpolTMD}) and (\ref{protontransversity}) to evaluate the azimuthal asymmetry. It should be noted that in our calculations, we neglect the contributions from sea quarks.
	The covered kinematical ranges of the COMPASS experiment are given by~\cite{COMPASS:2017jbv} 
	\begin{align}
		&0.05<x_N<0.4,\quad 0.05<x_\pi<0.9, \quad -0.3<x_F<1 \nonumber\\&
		4.3\ \mathrm{GeV}<Q<8.5\ \mathrm{GeV},
        \quad s=357~\mathrm{GeV}^2 .
		\label{eq:kinematics}
	\end{align}

 	\begin{figure}
		\includegraphics[scale=0.52]{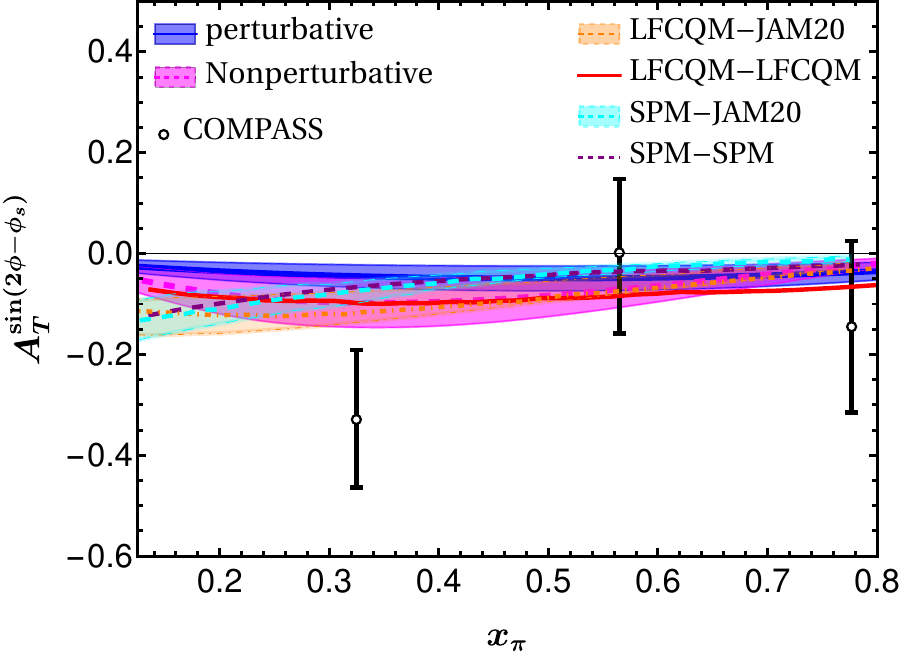}
		\includegraphics[scale=0.52]{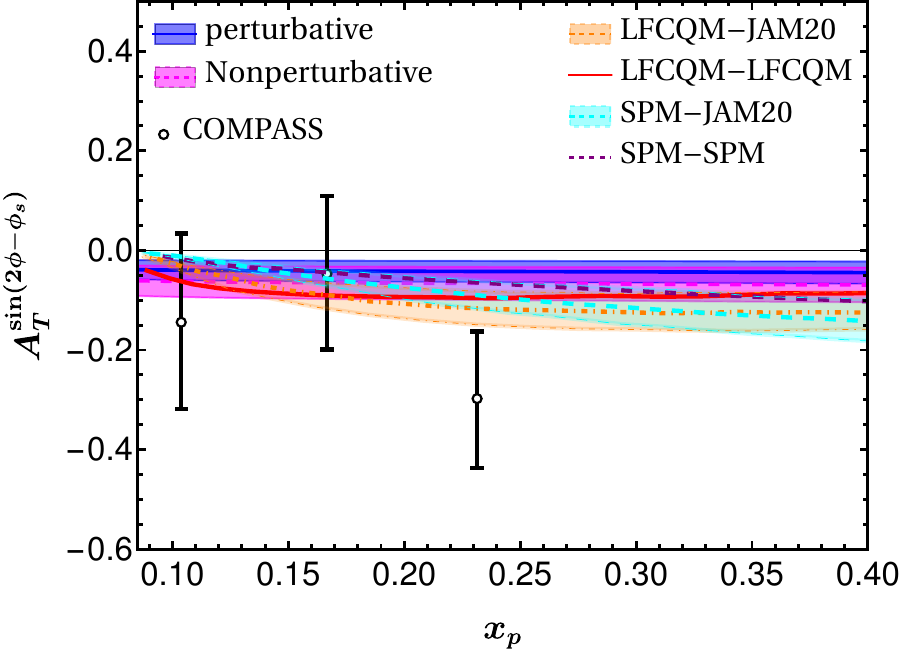}\\
		\includegraphics[scale=0.52]{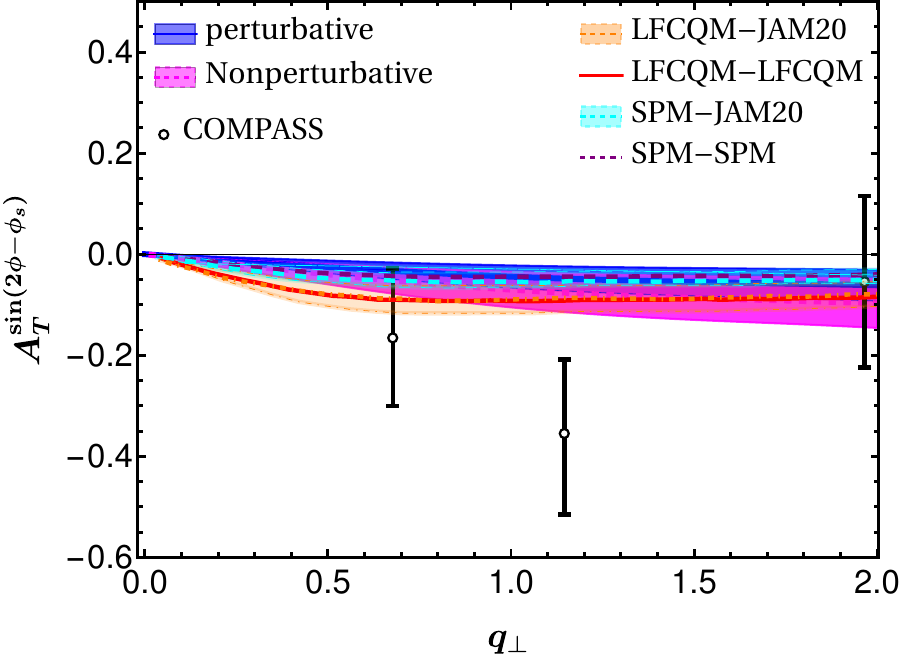}
		\caption{The $\sin(2\phi-\phi_{s})$ azimuthal asymmetry from $\pi^{-}N^{\uparrow}$ Drell-Yan process. The panels from top to bottom show the variation of the azimuthal asymmetry with $x_{\pi}$, $x_{N}$, and $q_{\perp}$, 
 respectively. The black open circles represent the COMPASS data~\cite{COMPASS:2017jbv}. Our estimations (blue and magenta bands) are compared with the results obtained  from pure model and hybrid calculations reported  in Ref.~\cite{Bastami:2020asv}. The orange (LFCQM-JAM20) and cyan (SPM-JAM20) bands represnt the hybrid computations, where the nonperturbative input for $h_{1,\pi^-}^{\perp q}$ is taken from the LFCQM and the SPM, respectively but the proton transversity TMD $h_{1,p}^{q}$ is adopted from the JAM20 global fit. The red-solid (LFCQM-LFCQM) and purple-dashed (SPM-SPM) lines correspond to the pure-model calculations, where both $h_{1,\pi^-}^{\perp q}$ and $h_{1,p}^{q}$ are taken from the LFCQM and the SPM, respectively. }
		\label{fig:asymmetry}
	\end{figure}
	
Based on the TMD factorization formalism as stated in Eqs.~(\ref{eq:asymmetry}), (\ref{eq:FUU}), and (\ref{eq:FUT}), we present our numerical results of the $\sin(2\phi-\phi_{s})$ azimuthal asymmetry in the pion-induced Drell-Yan process in Fig.~\ref{fig:asymmetry}, where we compare our
predictions with the COMPASS data~\cite{COMPASS:2017jbv} in the kinematical region given in Eq.~\eqref{eq:kinematics}.  We evolve our pion TMDs from the
model scale $Q_i\sim 0.316$ GeV~\cite{Ahmady:2018muv} to the scale $Q_f\sim 6.4$ GeV relevant to the
experimental data for the asymmetries following QCD
evolutions discussed in section~\ref{TMDevolsec}. Meanwhile, the proton TMDs are also evolved from the model scale  of the quark-diquark model $Q_i\sim 0.8$ GeV~\cite{Maji:2017bcz} to the relevant experimental scale. We perform the integration over transverse momentum $q_{\perp}$ in the $0<q_{\perp}<2$ GeV range that validates the TMD factorization in the $q_{\perp}\ll Q$ kinematic region~\cite{Sun:2013hua}.  The lines in Fig.~\ref{fig:asymmetry} represent the results calculated from the center values of the model parameters, while the bands represent the uncertainties in our model calculations determined by the uncertainties of those parameters~\cite{Ahmady:2018muv,Maji:2017bcz}. In this figure, the black circles show the experimental data measured by the COMPASS Collaboration~\cite{COMPASS:2017jbv} with the error bars
	corresponding to the sum of the systematic error and the
	statistical error. From the  top to bottom panels of the figure show the asymmetry as a  functions of $x_{\pi}$, $x_{p}$, and $q_{\perp}$, respectively. Neglecting the sea quark contributions, $A_{T}^{\sin (2\phi-\phi_s)}\propto - h_{1,\pi^-}^{\perp(1)\bar{u}} (x_\pi) h_{1,p}^{u}(x_p)$. Note that both $h_{1,\pi^-}^{\perp(1)\bar{u}}$ and $h_{1,p}^{u}(x_p)$ are positive in our model calculations, as can be seen from Figs.~\ref{fig:BMTMDmoments} and \ref{fig:quarkTransversity}, respectively. Consiquently, Fig.~\ref{fig:asymmetry} displays a negative $\sin(2\phi-\phi_{s})$ azimuthal asymmetry in the $\pi^{-}$-p Drell-Yan obtained from our pure model calculations, which is compatible with the COMPASS data.  Based on the analysis of the COMPASS data~\cite{COMPASS:2017jbv}, it can be inferred that the sign of pion Boer-Mulders function is positive. This is an important observation, which can be used to test the process dependence of other chiral-odd functions.

 We illustrate
the differences between the asymmetries generated by using the perturbative and nonperturbative gluon rescattering kernels~\cite{Ahmady:2019yvo} for the pion Boer-Mulders TMD. We find that both the perturbatively and nonperturbatively generated asymmetries are more or less consistent with the experimental data. It can also be noted that the nonperturbatively generated asymmetries are slightly larger in magnitude compared to that for the the perturbatively  generated asymmetries.

In Fig.~\ref{fig:asymmetry}, we also compare our predictions for the $\sin(2\phi-\phi_{s})$ asymmetry with the results reported in Ref.~\cite{Bastami:2020asv}, where the nonperturbative input for the pion Boer Mulders TMD
is taken from the LFCQM model~\cite{Pasquini:2014ppa} and the SPM~\cite{Gamberg:2009uk}, and the proton transversity TMD is adopted from the LFCQM~\cite{Pasquini:2008ax,Boffi:2009sh,Pasquini:2011tk} and the SPM~\cite{Gamberg:2007wm} as well as from the available parametrizations of TMDs extracted from the experimental data by JAM20 Collaboration~\cite{Cammarota:2020qcw} and Torino Collaboration~\cite{Anselmino:2013vqa}. We find that our results are compatible with the predictions yiedling from both the pure-model and the hybrid calculations~~\cite{Bastami:2020asv}. This allows us to quantitatively assess the holographic light-front QCD models in
future when more precise data will become available.

	\section{conclusion}\label{concl}
	
	In this work, we studied the $\sin(2\phi-\phi_{S})$ azimuthal asymmetry in the single transversely polarized $\pi^-p$ Drell-Yan process with focus on the kinematics of the COMPASS experiment.
	The asymmetry originates from the convolution of the Boer-Mulders function of the pion beam and the transversity TMD of the proton target. As no phenomenological extractions
are available for the pion Boer-Mulders TMD, we employed the holographic light-front QCD model for the pion, which leads to an excellent simultaneous description of a wide
class of different and related pion observables together with widely used  quark-diquark model for the proton. 
The gluon rescattering is crucial to obtain nonzero pion's Boer-Mulders TMD. We investigated the use of a
nonperturbative SU$(3)$ gluon rescattering kernel going beyond the usual approximation of perturbative U$(1)$ gluons. After implementing the TMD evolution effect, we found fair agreement between the first moment of the pion's Boer-Mulders function generated by both the perturbative and the nonperturbative gluon rescattering kernels and the COMPASS data. Meanwhile,  the transversity distributions of the proton computed in the quark-diquark model and the Bacchetta-Radici global fits~\cite{Radici:2018iag} displayed good mutual agreement, however, the quark-diquark model predictions were somewhat underestimated compared to the other global analyses and model predictions.
	
	We then presented the pure-model computations of the  $\sin (2\phi-\phi_S)$ azimuthal asymmetry at the kinematics of COMPASS.
	Our analysis showed that the $\sin(2\phi-\phi_{S})$ asymmetry at COMPASS can be qualitatively described (sign and magnitude) by the present analysis of the TMDs of the pion within the framework of the holographic light-front QCD and the proton TMDs in a light-front quark-diquark model constructed by the soft-wall AdS/QCD. 
 In regard to the interpretation of the first data from the pion-induced Drell-Yan process with polarized protons, we observed a robust picture.  The data favor a positive quark Boer-Mulders distribution in the pion. More precise upcomig data
from COMPASS and other experimental facilities will allow us to solidify the picture. Our investigation helped to provide quantitative tests of the application of holographic light-front QCD model to the description of pion.

	\section*{Acknowledgments}
	We would like to thank Dipankar Chakrabarti for fruitful discussions. The work of CM is supported by new faculty start up funding by the Institute of Modern Physics, Chinese Academy of Sciences, Grant No. E129952YR0.  CM also thanks the Chinese Academy of Sciences Presidents International Fellowship Initiative for the support via Grants No. 2021PM0023. 
	\bibliography{References.bib}
\end{document}